%% file: main.tex
\documentclass[10pt,conference, final]{IEEEtran}
\IEEEoverridecommandlockouts
\usepackage{amsmath,amssymb,amsfonts}
\usepackage{algorithmic}
\usepackage{graphicx}
\usepackage{textcomp}
\usepackage{xcolor}
\def\BibTeX{{\rm B\kern-.05em{\sc i\kern-.025em b}\kern-.08em
    T\kern-.1667em\lower.7ex\hbox{E}\kern-.125emX}}
    
\usepackage{lineno}
\usepackage[utf8]{inputenc}
\usepackage[T1]{fontenc}
\usepackage{xspace}
\usepackage{amssymb}
\usepackage{amsmath}
\usepackage{amsfonts}
\usepackage{amsthm}
\usepackage{logicproof}
\usepackage{lscape}
\usepackage{url}
\usepackage{multirow}
\usepackage{array}
\usepackage{paralist}

\usepackage{listings}
\RequirePackage{etex}
\usepackage{etex}
 
\usepackage{tikz}
\usetikzlibrary{arrows}
\usetikzlibrary{shapes}
\usetikzlibrary{patterns}
\usetikzlibrary{positioning}

\usepackage[colorinlistoftodos,prependcaption]{todonotes}
\newboolean{showcomments}
\setboolean{showcomments}{true}
\ifthenelse{\boolean{showcomments}}
 { \newcommand{\mynote}[2]{
      \fbox{\bfseries\sffamily\scriptsize#1}
        {\small$\blacktriangleright$\textsf{\textcolor{red}{{\em #2}\bf }}$\blacktriangleleft$}}}
        { \newcommand{\mynote}[2]{}}
\usepackage{tabularx}
\newcolumntype{C}{>{\centering\arraybackslash}X}
\newcolumntype{R}{>{\raggedleft\arraybackslash}X}

\usepackage{booktabs}

\definecolor{mymauve}{rgb}{0.58,0,0.82}
\definecolor{mygrey}{rgb}{0.43, 0.5, 0.5}

\usepackage{pbox}

\makeatletter
\newcommand*{\centerfloat}{%
  \parindent \z@
  \leftskip \z@ \@plus 1fil \@minus \textwidth
  \rightskip\leftskip
  \parfillskip \z@skip}
\makeatother 

\input{utils.tex}

\usepackage{cite}

\usepackage{etex}
\usepackage{tikz}
\usetikzlibrary{arrows.meta}
\usetikzlibrary{arrows}
\usetikzlibrary{shapes}
\usetikzlibrary{patterns} 
\usetikzlibrary{positioning}
\usetikzlibrary{calc}

\usepackage{filecontents}

\begin{document}

\title{The Maven Dependency Graph: a Temporal Graph-based Representation of Maven Central
\thanks{This work has been  partially supported by the STAMP ICT-16-10 No.731529, by the Wallenberg Autonomous Systems and software Program (WASP), and by Orange.}
}

\author{

\IEEEauthorblockN{Amine Benelallam\IEEEauthorrefmark{1}, Nicolas Harrand\IEEEauthorrefmark{2}, C\'esar Soto Valero\IEEEauthorrefmark{2}, Benoit Baudry\IEEEauthorrefmark{2}, Olivier Barais\IEEEauthorrefmark{1}}
\IEEEauthorblockA{
\IEEEauthorrefmark{1}\textit{Univ Rennes, Inria, CNRS, IRISA, Rennes, France}, firstname.lastname@inria.fr \\
\IEEEauthorrefmark{2}\textit{KTH Royal Institute of Technology, Stockholm, Sweden}, lastname@kth.se }

}
\maketitle

\begin{abstract}
The Maven Central Repository provides an extraordinary source of data to understand complex architecture and evolution phenomena among Java applications. As of September 6, 2018, this repository includes 2.8M artifacts (compiled piece of code implemented in a JVM-based language), each of which is characterized with metadata such as exact version, date of upload and list of dependencies towards other artifacts. 

Today, one who wants to analyze the complete ecosystem of Maven artifacts and their dependencies faces two key challenges: (i) this is a huge data set; and (ii) dependency relationships among artifacts are not modeled explicitly and cannot be queried. In this paper, we present the \emph{Maven Dependency Graph}. This open source data set provides two contributions: a snapshot of the whole Maven Central taken on September 6, 2018, stored in a graph database in which we explicitly model all dependencies; an open source infrastructure to query this huge dataset.
\end{abstract}


\section{Introduction}
\label{sec:intro}
\input{intro.tex}

\section{Description of the dataset}
\label{sec:dataset}
\input{dataset.tex}

\section{Research opportunities}
\label{sec:research}
\input{research.tex}

\section {Limitations \& threats to validity}
\label{sec:limitations}
\input{limitations.tex}

\section{Conclusion}
\label{sec:conclusion}
\input{conclusion.tex}

\bibliography{biblio}
\bibliographystyle{plain}
\end{document}

%% file: intro.tex
 

Maven Central is one of the most popular and widely used repositories of JVM-based artifacts. It stores a large collection of software binaries together with their corresponding metadata in a well-defined structure, characterizing the exact version, date of upload and list of dependencies towards other artifacts. Preaching for reusability and ease of dependency management since its launch in 2004, Maven Central keeps attracting open-source projects and software vendors, reaching nowadays\footnote{September 6, 2018} more than $2.8M$ unique artifacts.  

Maven Central holds a treasure-worth big data that can reveal valuable insights about software engineering processes, evolution, and trends thanks to recent advances in big data analysis techniques. However, it is currently extremely challenging to perform analyses at the scale of the whole Maven Central. First,
dependency relationships among artifacts are not modeled explicitly and cannot be queried. This data should be made available in a format that is conveniently consumed by big data processing and analysis frameworks, to run  Empirical Software Engineering studies. Second, exporting all data from Maven Central is highly time and resource consuming because of the huge number of artifacts.

In this work, we showcase the \mdg, a novel dataset that aims at letting the Software Engineering community run empirical studies on the whole Maven Central. This open source graph\footnote{\url{https://zenodo.org/record/1489120}} includes metadata about $2,4M$ Maven Central artifacts, indexed by deployment date in the Gregorian calendar. The graph includes more than $9M$ explicit dependencies between artifacts as well as other relationships to represent artifacts' version precedence.  Artifacts are described by the 3-tuple `GroupId:ArtifactId:Version', distinguishing different versions of a given library (`GroupId:ArtifactId')\footnote{Throughout the rest of the paper, we use library to refer to the couple \textit{GroupId:ArtifactId}}. This represents 85\% of all Maven artifacts and their dependencies, as of September 6, 2018.

Our second contribution comes in the form of Maven-graph procedures. These procedures aim at facilitating queries over the big \emph{Maven Dependency Graph}. This collection of procedures implement common queries; such as artifacts retrieval in time or per version-range, and many other features. We provide a custom Neo4j~\cite{neo4j} Docker image shipping the entire dataset, together with the procedures plugin. 
These procedures, as well as our Maven Miner tool that can collect a snapshot of Maven Central and store it into a graph database, are open-source and available online~\cite{maven-miner}.

The \mdg is intended to answer high-level research questions about artifacts releases, evolution, and usage trends over time. It also provides a solid basis to select relevant subsets of artifacts for assessing specific software engineering challenges. The queries over the \mdg can range from pattern matching techniques, \eg, `\textit{How often do libraries release new versions?}',  to advanced big data analysis, such as `\textit{What are the most influential artifacts in the Maven Central?}' or, even predictive models using machine learning, \eg, `\textit{What artifacts are more likely to be adopted or overlooked by the community?}'.
 
The \mdg is related to the Maven Dependency Dataset~\cite{mdd} (MDD). This previous dataset captured a snapshot of the Maven Central on July 30, 2011 and aimed at supporting large-scale research on libraries' releases and dependencies. Since then, the Maven Central has $13.5\times$ more artifacts and $14.7\times$ more dependencies. Hence, we believe that an updated dataset is valuable for the software engineering community. Yet, because of this huge growth, our dataset resolves dependencies only at the artifacts level, by opposition to MDD that abstracts dependencies at the source code level too.\looseness=-1


%% file: dataset.tex
In this section, we provide a general overview of the dataset. First, we describe the data schema. Later we present the data retrieval methodology and tooling.

\subsection{Overview \& Schema}
\begin{figure}[ht]
\centering
  \resizebox{0.9\linewidth}{!}{  
	  \input{schema.tex}
  }
  \caption{Maven dependency graph schema}
  \label{fig:schema}
\end{figure}
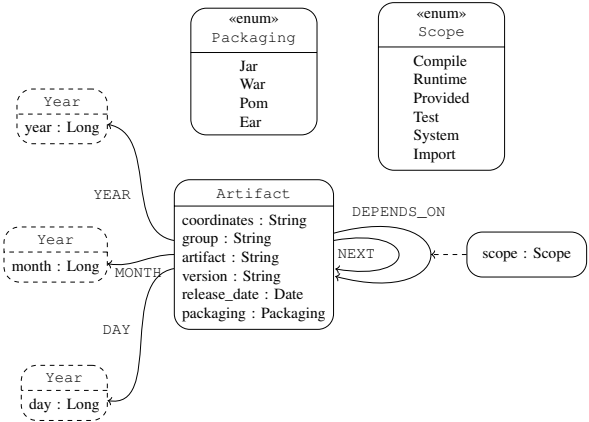

We rely on a temporal graph-based representation to capture the artifacts' dependency graph of the Maven Central. Figure~\ref{fig:schema} shows a simplified schema of the \mdg. 
Formally,  $\mathcal{M}$, the \mdg, is defined as a tuple $(\mathcal{A},\mathcal{C},\mathcal{D}, \mathcal{N})$.  $\mathcal{A}$ is a set of nodes that model the Maven artifacts. Every artifact node has a timestamp referring to its deployment date.  
Each node holds a set of properties: its \textit{groupId}, \textit{artifactId}, \textit{version}, and \textit{packaging}.
The property \textit{coordinates} is used to identify artifact nodes uniquely. Its value comes in the form 'group:artifact:version`.
$\mathcal{C}$ are calendar nodes, represented by dashed boxes in Figure~\ref{fig:schema}. They operate as a proxy to artifacts timestamp release date property. Their main intent is to temporally index the artifacts by their release date. 
$\mathcal{D}$ is a set of dependency relationships. Every $d \in \mathcal{D}$ can be regarded as a couple ($a_i,a_i$) $\in \mathcal{A} \times \mathcal{A}$  where $a_i$ and $a_j$ are respectively the user and provider of a library. A dependency $d$ has a scope, which limits the transitivity of a dependency. $\mathcal{D}$ is depicted by the label \texttt{DEPENDS\_ON} in Figure~\ref{fig:schema}. Finally, $\mathcal{N}$ is the set of version precedence relationships, represented by the label \texttt{NEXT}. Every $n \in \mathcal{N}$ is described as a couple ($n_i$,$n_j$) $\in \mathcal{A} \times \mathcal{A}$ where $n_i$ and $n_j$ are respectively a given artifact and its next release.


\begin{logicproof}{1}
    \forall i,j \in \mathcal{A}, coord(i) = coord(j) \implies i = j \label{eq:1}\\
    \forall i \in \mathcal{A}, coord(i) \neq \emptyset \label{eq:2}\\
    \forall d \in \mathcal{D}, scope(d) \neq \emptyset\label{eq:3}
\end{logicproof}

Our schema adheres to a set of constraints, namely uniqueness and existence. Constraints~\eqref{eq:1}~\eqref{eq:2}~\eqref{eq:3} depict few of them. The first constraint assures that nodes are uniquely identified by their coordinates. Whilst, the remaining constraints assure that every resolved artifact contains some mandatory properties. Other constraints such as uniqueness of edges and calendar nodes are not covered in this paper.
\subsection{Descriptive statistics}

The \mdg represents a snapshot of the Maven Central Repository from September 6, 2018. Descriptive statistics can be found in Tables~\ref{tab:stat}~and~\ref{tab:perc}.

While the Maven Central index contains $\sim3,2$M artifacts, almost $400,000$ entries are duplicated, leaving us with 2.8M artifacts. We retrieved metadata and dependency information for $2,407,335$ artifacts identified by their unique coordinates in the form of \textit{GroupId:ArtifactId:Version}. The missing artifacts are either deployed in another artifact repository or their \texttt{pom.xml} is corrupted. As shown in Table~\ref{tab:stat}, these artifacts belong to $35,699$ unique groups and represent $223,478$ libraries (i.e., collections of artifacts with different versions but similar groupId and artifactId). Libraries exist in  $\sim10$ versions on average, with a minimum of 1 version and a maximum of $2,182$ versions, totaling $2,183,845$ upgrade operations. Other percentile values of versions count are provided in Table~\ref{tab:perc}.

The \mdg has $9,715,669$ edges, i.e., directed dependency relationships, regardless of their dependency scope. The graph has a density of $\sim4$. We call outgoing edges \texttt{dependencies}, while incoming edges are \texttt{usages}. Table~\ref{tab:perc} provides the minimum and maximum values of usages and dependencies, as well as the percentile values.

\begin{table}
    \centering
	\caption{Descriptive statistics about the graph of Maven libraries collected from Maven central in September 6, 2018}
	\label{tab:stat}
	\begin{tabular}{lr}
		\hline
		Property & Number \\
		\hline
		Total artifacts & 2,407,335 \\ 
		 Libraries & 223,478\\ 
		 Groups & 35,699 \\
		 Upgrades  & 2,183,845\\
		 Dependency relationships & 9,715,669 \\
		 Density & 4.03 \\ 
		\hline
	\end{tabular}
\end{table}
\begin{table}
    \caption{Percentile values of the Maven dependency graph relationships}
    \centering
    \begin{tabular}{lccccr}
    \hline 
     & \multicolumn{3}{c}{Percentiles} \\
    \cline{2-4}
    & 25th & 50th & 75th & min & max\\
    \hline
    Dependencies & 2 & 30 & 89 & 0 & 402 \\
    Usages & 0 & 38 & 405 & 0 & 47,951 \\
    Versions & 2 & 110 & 485 & 1 & 2,182 \\
    \hline
    \end{tabular}
    \label{tab:perc}
\end{table}

\subsection{Data Collection Methodology \& Tooling}

\begin{figure}[ht]
\centering
 \includegraphics[width=.9\linewidth]{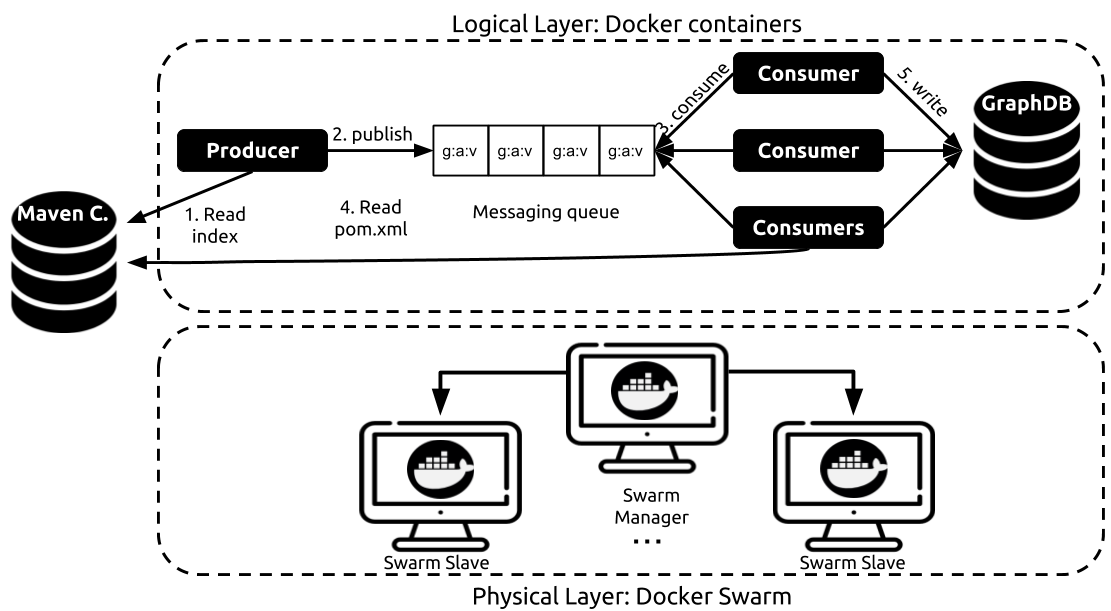}
  \caption{Maven dependency graph schema}
  \label{fig:architecture}
\end{figure}

Data collection involved retrieving  \textit{pom.xml} files (at least one per artifact) from Maven Central, parsing them to retrieve metadata, and finally storing this information into a graph database. This a time-consuming process that we distributed  on top of a Docker Swarm cluster.
Figure~\ref{fig:architecture} shows the overall architecture and methodology. The process  ran on a cluster of 4 identical machines running an Ubuntu 18.04 LTS. Each machine has 16 Gb of RAM and 4 identical CPUs (MD A10-7700K APU with Radeon(TM) R7 Graphics, 2.105 GHz). One machine played the role of a Swarm Master while the others were Swarm Slaves.

We rely on a producer-consumer pattern to distribute the computation (upper part of Figure~\ref{fig:architecture}). The producer is responsible for reading the Maven Central Index, wrapping artifact coordinates into messages, and publishing them in a shared messaging queue. On the other side, each consumer retrieves one artifact coordinate at a time. For each artifact, the consumer resolves the artifact's meta-data as well as its direct dependencies and store them in a graph database. Finally, the consumer acknowledges the broker having finished processing the message. In case of a consumer failure, the message broker puts back the message in the queue. Note messages are removed from the queue only if the corresponding consumer acknowledges so. Moreover, a message is processed by only one consumer at a time. When all artifacts are resolved, a post-processing phase is responsible for creating artifacts versions chains.\looseness=-1

For message queuing, we use RabbitMQ~\cite{rabbitmq}, a scalable and widely used message broker.
As for the graph database backend, we rely on Neo4j~\cite{neo4j}, one of the most popular NoSQL databases. 
It comes along with a powerful SQL-like graph query language, Cypher~\cite{cypher}. This simplifies the exploitation of the dataset in a very simple manner. Finally, to fetch artifacts from Maven Central and resolve their direct dependencies, we use Aether~\cite{aether} Eclipse, a Java library to manage artifact repositories. 
  
\section{The Maven Dependency Graph in Action}
\label{sec:action}
We have implemented a graph-based persistence backend for the \mdg. This allows interested users to exploit the dataset through the Neo4j web interface, leveraging the Cypher graph querying language~\cite{cypher}. Cypher is an open-source declarative language to specify graph queries with patterns. Multiple drivers have been implemented around Cypher, allowing its integration in other graph databases, such as SAP HANA, or distributed processing frameworks like Spark and Hadoop~\cite{cypher-usage}.

To further simplify queries on \mdg, we leverage Cypher procedures and functions. This mechanism supports the extension of Cypher by writing custom code, deploying it into the database, and calling it from Cypher. We have implemented a set of functions and procedures to simplify the description of queries involving versions comparison, artifacts selection by versions' range or by date. Listing~\ref{lst:query-dep} shows a usage example of such functions. A complete list of Maven-graph procedures can be found online~\cite{miner-proc}.

In the following, we illustrate some usage examples.  

\paragraph*{The artifacts deployed in 2018} Listing~\ref{lst:query-temp} lists all the artifacts $n$ that have been deployed during $2018$ and use `Junit', regardless of the scope.  

\begin{lstlisting}[language=SQL,
                   caption={Cypher query computing the number of artifacts deployed in 2018},
                   label={lst:query-temp},
                   morekeywords={MATCH,CALL,CONTAINS,DEPENDS_ON, NEXT, ASSERT, IS, WITH, RETURN }]
MATCH (y:Calendar {year : 2018})<-[:YEAR ]-(n:Artifact)-[r:DEPENDS_ON]->(:junit)
RETURN n.coordinates AS coordinates
\end{lstlisting}

\paragraph*{Number of versions per library} This example shows how to make use of the precedence relationship (\texttt{NEXT}) to compute the number of versions per library. Listing~\ref{lst:query-version} depicts the corresponding Cypher query. The query runs in two steps. Given a node with no incoming edges, it selects the longest path of the \texttt{NEXT} relationship and returns its length, together with the nodes' groupId and artifactId. The second step simply selects nodes with neither incoming nor outgoing \texttt{Next} relationship and return 1 (\ie one version) together with the groupId and artifactId. The results of the two steps are aggregated using the  \textbf{\texttt{UNION}} operation.

\begin{lstlisting}[language=SQL,
                   caption={Cypher query computing the number of versions per library},
                   label={lst:query-version},
                   morekeywords={MATCH,CALL,CONTAINS,DEPENDS_ON, NEXT, ASSERT, IS, WITH, RETURN }]
MATCH (n:Artifact) WHERE NOT (n)<-[:NEXT]-() WITH n
MATCH p=(n)-[:NEXT*]->(m) WHERE NOT (m)-[:NEXT]->()
RETURN n.groupID as groupId, n.artifact AS artifactId, length(p) AS versions
UNION
MATCH(n:Artifact) WHERE NOT (n)<-[:NEXT]-()  AND NOT  ()<-[:NEXT]-(n)
RETURN n.groupID AS groupId, n.artifact AS artifactId, 1 AS versions                  
\end{lstlisting}

\paragraph*{Artifacts using older `JUnit` versions compared to libraries they are using} The query in Listing~\ref{lst:query-dep} simply selects all the nodes $n$ and $m$ where $n$ depends on $m$, only on the `Test' scope, but $n$ uses an older version of JUnit than $m$.  We use our custom procedure $maven.miner.version.isLower$ to check versions precedence. It takes as parameters an artifact node and a version as a String and returns true if the node's version is strictly older than the given version. We use the label `junit' instead of `Artifact' to avoid checks on the groupId value and speed up query execution, relying on labels indexes.

\begin{lstlisting}[language=SQL,
                   caption={Cypher query computing artifacts using older `JUnit` versions},
                   label={lst:query-dep},
                   morekeywords={MATCH,CALL,CONTAINS,DEPENDS_ON, NEXT, ASSERT, IS, WITH, RETURN }]
MATCH (j1:junit)<-[:DEPENDS_ON {scope : "Test"} {}]-(n:Artifact)-[:DEPENDS_ON]->(m:Artifact)-[:DEPENDS_ON  {scope : "Test"}]->(j2:junit)
WHERE maven.miner.version.isLower(j1,j2.coordinates) 
RETURN n.coordinates AS source, m.coordinates AS target                 
\end{lstlisting}

%% file: schema.tex
\begin{tikzpicture}[
  ->,
    simplenode/.style={
      draw,
      rectangle,
      rounded corners=10pt,
      inner sep = 10pt
    },
    node/.style={
      draw,
      node distance=6cm,
      rectangle split,
      rectangle split parts=2,
      rounded corners=10pt,
      inner sep = 5pt
    }
  ]
    \node[node] (art) {
        \texttt{Artifact}
        \nodepart{second}
        \pbox{3.5cm} {
         coordinates : String \\
         group : String\\
         artifact : String\\
         version : String\\
         release\_date : Date\\
         packaging : Packaging
        }
    };
      \node[node, above right = .2cm and 1cm of art] (scope) {
      \parbox{2.5cm}{\centering
      <<enum>>\\
      \texttt{Scope}}
        \nodepart{second}
        \pbox{2.5cm} {
         Compile\\
         Runtime \\
         Provided \\
         Test \\
         System \\
         Import
        }
    };
       \node[node, above = 1cm of art] (packaging) {
         \parbox{2.5cm}{\centering
            <<enum>>\\
      \texttt{Packaging}}
        \nodepart{second}
        \pbox{1.5cm} {
        Jar \\
        War \\
        Pom \\
        Ear 
        }
    };
    \node[node, dashed, above left = .8cm and 1.5cm of art] (year) {
        \texttt{Year}
        \nodepart{second}
        year : Long
    }; 
    \node[node,dashed,  left = 1.5cm of art] (month) {
        \texttt{Year}
        \nodepart{second}
        month : Long
    }; 
    \node[node,dashed, below left = .8cm and 1.5cm of art ] (day) {
        \texttt{Year}
        \nodepart{second}
        day : Long
    };
    \node[simplenode, right = 3cm of art] (scopeAss) {scope : Scope}; 
    \node [right = 1.95cm of art](dummy){};

    \path[->](art) edge [in=-15,out=15,loop, align=right] node [above right = .7cm and -1.9cm of art] {\texttt{DEPENDS\_ON}} (art);
    \path[->](art) edge [in=-10,out=10,loop, align=right] node [above right = -.2cm and -1.5cm of art] {\texttt{NEXT}} (art);
    \path[->](art) edge [in=-10,out=170, align=left] node [above left = -.5cm and .1cm of art] {\texttt{YEAR}} (year);
    \path[->](art) edge [in=-10,out=180, align=left] node [ above left = -.5cm and -.6cm of art] {\texttt{MONTH}} (month);
    \path[->](art) edge [in=-10,out=190, align=left] node [below left = -.5cm and .1cm of art] {\texttt{DAY}} (day);
    \draw[dashed](scopeAss)--(dummy);
  \end{tikzpicture}

%% file: research.tex
In this section, different types of empirical analyses that can leverage our dataset, as well as some research opportunities it can open up.

\begin{enumerate}
     \item \textbf{Libraries maintenance:} Java Libraries continuously evolve by releasing new versions with new functionality or enhanced performance. However, in some cases, clients decide not to upgrade their dependencies to newer versions. As a result, library maintainers may decide to continue maintaining parallel versions. When does this phenomenon happen? When do project maintainers decide to maintain two parallel versions? Why? Who are the clients that stick with an older version? To answer these questions, we should first identify libraries that keep on maintaining multiple versions. The \mdg can help to identify these projects,  by comparing versions precedence of artifacts and crossing them with their release date. Subsequently, we can identify the clients that keep using older versions.  Another side-effect to libraries evolution is the growing complexity of latest releases. When facing such issues, libraries' maintainers decide to decompose the library into different ones, ending this library's lifetime. Another interesting point could also be detecting two or more artifacts merging into a single one. The \mdg supports the empirical inquiry of this kind of behavior. 
    \item \textbf{Libraries adoption trends: Wisdom of the crowd Vs. Hype-driven development Vs. legacy} This question focuses on end-users instead of library maintainers.  What are the motivations that steer their decision to use a specific library? Do they behave according to Rogers’ theory~\cite{rogers2003diffusion} of Diffusion of innovation? Are there any organizational or social factors influencing these decisions? The wisdom of the crowd principle favors the collective opinion of a group of individuals rather than that of a single expert. It has been used as a form of crowd-sourcing in software engineering for numerous tasks~\cite{mao2017survey}. In particular, Mileva~\etal~\cite{Mileva09} encourages the wisdom of the crowds as a principle to assess developers deciding which library versions to use, and thus, avoiding some pitfalls experienced by other developers. However, many lead developers have been warning about the doom this might bring to their products. This anti-pattern development is called Hype-Driven Development. A more recent work~\etal~\cite{gai17modeling} leverages the same principle  to recommend consented library updates. Their recommendation system relies on a graph that is very similar to our dataset. To evaluate their approach, they constructed a graph containing $188,951$ nodes of $6,374$ maven unique artifacts. We believe that a replication with the \mdg that is 3 orders of magnitude larger would improve the quality of such recommenders.
    \item \textbf{House of cards Vs. sustainable software:} In 2016, Sonatype analyzed $25,000$ applications and showed that $6.8\%$ had at least one security flaw tied to the use of a vulnerable library ~\cite{sonatype-report}. What are these libraries that once are vulnerable? How much vulnerable are these clients? A recent work~\cite{Kikas17} attempted to answer a similar question by studying the state of dependency update practices and the structure of dependency networks in JavaScript projects. Pashchenko and colleagues~\cite{pashchenko18} introduced the concept of halted dependencies to describe the libraries that are no longer maintained. Together with other information extracted from code repositories, and using code-based analysis of patches, the authors were able to implement a methodology to assess developers quantifying the vulnerability of their tools when using 3rd-party libraries. We believe that the \mdg is perfectly fit to answer this kind of questions. 
\end{enumerate}

\section{Availability}
\label{sec:availability}
We used  Maven-miner, a set of tools and facility scripts, to collect the \mdg. 
The source code of Maven-miner is publicly available online~\cite{maven-miner}. Maven-miner runs in different setups, standalone, docker-compose mode, or docker-swarm mode. Ready to use Docker images and scripts are also available online. Instructions on how to the Maven-miner scripts can be found in the wiki section of the tool's repository. Note it is discouraged to use the standalone mode to resolve all the dependencies in the Maven Central as it may take months to finish. The standalone mode is only intended to resolve a small set of artifacts.

The \mdg  is publicly available and accessible from the tool's repository. 
For ease of use, it comes in the form of docker images shipped with all the facility procedures simplifying data exploitation. For use outside of the Neo4j ecosystem, we have also released CSV files. The entirety of data can be found online: 

\begin{center}
{\fbox{\url{https://zenodo.org/record/1489120}}}
\end{center}

%% file: limitations.tex
Due to some technical limitations, we were,
not able to resolve all the information about existing artifacts in the Maven Central. In particular, we do not consider artifacts dependencies that are hosted outside of the Maven Central repository. For this reason, some metrics like libraries usage and dependencies may not reflect the reality.
Moreover,  our dataset lacks some low-level information such as excluded dependencies. Consequently, querying the dependency tree of a given artifact may result in a super-set, including conflicting dependencies. \looseness=-1

Although the proposed schema was designed to improve queries execution, very complex queries involving computation expensive operations, such as transitive usages traversals, require a significant computation power.
Finally, our collection is limited to the Maven Central repository, any findings reflect only the state of practice in the Maven repository, and it should not be generalized.\looseness=-1

%% file: conclusion.tex
We presented the \mdg, an open-source dataset that aims at enabling the Software Engineering community to conduct large-scale empirical studies on Maven Central. To ease the exploitation of this dataset, we provide a custom Neo4j Docker image shipping the entire dataset. It comes along with a very large collection of procedures implementing common graph queries and utility functions.  We also introduced Maven Miner, a set of tools that enable the collection of the \mdg. Both \mdg and Maven miner are open-source and publicly available online.